\documentclass[pra,superscriptaddress,showpacs,twocolumn]{revtex4}
\usepackage{amsfonts}
\usepackage{amsmath}
\usepackage{amssymb}
\usepackage{graphicx}

\newcommand{\ket}[1]{|#1\rangle}
\newcommand{\bra}[1]{\langle#1|}

\begin{document}
\title{Non-destructive cavity QED probe of Bloch oscillations in a gas of ultracold atoms}
\author{B.~M.~Peden}
\affiliation{JILA, NIST, and Department of Physics, University of Colorado at Boulder, CO, USA}
\author{D.~Meiser}
\affiliation{JILA, NIST, and Department of Physics, University of Colorado at Boulder, CO, USA}
\author{M.~L.~Chiofalo}
\affiliation{Department of Mathematics and INFN, University of Pisa, Pisa, Italy}
\author{M.~J.~Holland}
\affiliation{JILA, NIST, and Department of Physics, University of Colorado at Boulder, CO, USA}

\begin{abstract}
We describe a scheme for probing a gas of ultracold atoms trapped in an
optical lattice and moving in the presence of an external potential. The probe
is non-destructive and uses the existing lattice fields as the measurement device.
Two counter-propagating
cavity fields simultaneously set up a conservative lattice potential and a weak quantum
probe of the atomic motion. Balanced heterodyne detection of the probe field at the cavity
output along with integration in time and across the atomic cloud yield information about
the atomic dynamics in a single run. The scheme is applied to a measurement of the
Bloch oscillation frequency for atoms moving in the presence of the local gravitational
potential. Signal-to-noise ratios are estimated to be as high as $10^4$.
\end{abstract}
\date{\today}
\maketitle

\section{Introduction}

The simulation of many-body models using gases of ultracold atoms trapped in optical lattices
\cite{jaksch1998cba} has been successful in investigating many systems in condensed-matter
physics. Band physics in gases of non-interacting Fermi gases in periodic potentials has been
studied \cite{kohl2005fat}, quantum phase transitions such as the Mott insulator to superfluid
transition have been observed \cite{greiner2002qpt}, and strongly-correlated physics such as
in one-dimensional systems \cite{paredes2004tgg,stoferle2004tsi} has been investigated. In
these experiments, techniques such as time-of-flight measurements and Bragg spectroscopy
are typically employed to probe atomic  states and dynamics in optical lattices.

In this paper, we present an alternative method for optically probing atomic gases in optical
lattices subject to an external potential. The method is {\it in situ} and non-destructively
measures properties of the atomic motion via weak-coupling to the existing lattice fields. The
technique satisfies three main goals. The probe is weak so that the atoms can be continuously
monitored without affecting their dynamics; the existing lattice fields are employed as the probe,
so that no external interrogation fields are necessary; and the signal-to-noise-ratio (SNR) is
large enough for experimental detection. In a ring-cavity, two counter-propagating running-wave
modes interact with a gas of ultracold atoms and simultaneously set up both a conservative,
external lattice potential for the atoms and a weak, quantum optical probe of the atomic
center-of-mass dynamics. The probe field leaks out of the cavity and is detected with a
balanced heterodyne scheme at the cavity output.

This method is in a sense dual to strong measurement schemes such as time-of-flight
absorption imaging and Bragg spectroscopy. In these schemes, light from a strong source
is either absorbed by or scattered off of the atomic cloud. This allows for high resolution
images and a strong signal using only a single measurement, but the atomic sample
is destroyed in the process. Here, the probe field is very weak so that a continuous
measurement is made without affecting the atomic dynamics. Integration of the signal in
time and across the atomic cloud yields measurements of dynamical properties of the
atoms with a measurable SNR in a single experimental run at the price of losing information
about individual atoms and real-time dynamics.

The procedure is similar in nature to recent proposals for optical detection of many-body
atomic states. In one scheme, a weak probe beam is scattered off of atoms trapped in an
optical lattice into a cavity mode, and signatures of many-body states such as Mott
insulators and superfluids appear in the out-coupled fields~\cite{mekhov2007cel}. In
another, atoms in a lattice interact with two-counter-propagating ring-cavity modes, and
atomic number statistics can be inferred from the behavior of the cavity fields%
~\cite{chen2007cqd}.

Related techniques have been applied to nondestructive optical measurements of Rabi
oscillations in gases of Cs atoms~\cite{windpassinger2008npr}, of the Cs clock transition
pseudo-spin~\cite{chaudhury2006cnm}, and of nonlinear dynamics in cold gases%
~\cite{smith2004cwm}. In addition, state preparation such as atomic spin squeezing via
measurements on out-coupled cavity fields has been proposed~\cite{nielsen2008ass,%
meiser2008sso, mekhov2009qnm}. Finally, it has been demonstrated that the motion of
individual atoms in an optical cavity can be tracked by the transmission of a probe
field~\cite{hood2000acm}.

We here provide a test of the technique for the conceptually simple motion of non-%
interacting atoms in an optical lattice driven by a constant force, which leads to Bloch
oscillations~\cite{bendahan1996boa}. Besides its simplicity, this choice is motivated by
the fact that Bloch oscillations can be viewed as a general probe for investigating
quantum gases in optical lattices. These oscillations may be used in the
measurement of fundamental constants~\cite{clade2006dfs}, to provide levels of
precision up to $\delta g/g\approx10^{-7}$ in the measurement of the acceleration of
gravity~\cite{anderson1998mqi, roati2004ait, carusotto2005smf, ivanov2008cda}, and to
measure Casimir forces on small length scales~\cite{wolf2007fol}. When interactions
are significant, damping and destruction of Bloch oscillations provide information on
correlation-induced relaxation processes~\cite{freericks2008qbo, mehta2006ntq,%
oka2005gsd, buchleitner2003iid, dias2007fdb}. Finally, this investigation is a jumping-off
point for other optical measurement schemes, such as periodically driven lattices that act as a
spectroscopic probe of the atomic motion~\cite{ivanov2008cda}.

The paper is organized as follows. In Sec. \ref{sec:Model}, we present the details of the
system and detection scheme. In Sec. \ref{sec:Results}, we apply this scheme to the
detection of Bloch oscillations in an optical lattice. In Sec. \ref{sec:Conclusion}, we
summarize the main results of the paper and conclude with prospects for measurements
of many-body properties of gases of ultracold atoms trapped in optical lattices.

\section{Model and Detection Scheme}\label{sec:Model}

\begin{figure}[tb]
\includegraphics[width=80mm]{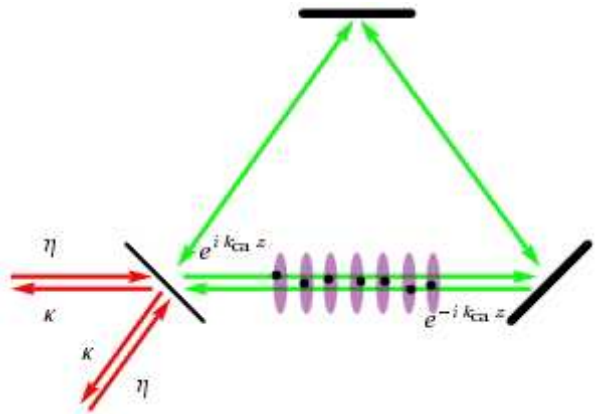}
\caption{(Color online.) Schematic of the coupled atom-cavity system. In-coupled
lasers set up two counter-propagating fields within the cavity. The atoms interact
with the cavity fields via the optical dipole potential.  Photons from the cavity beams exit
the cavity through the mirrors at a rate $\kappa$.}
\label{fig:systemschematic}
\end{figure}

Since the mathematical details of the system and detection scheme are somewhat
complicated, we briefly outline the physical basis of the
model and approximations used. This includes a discussion of obstacles in the way
of satisfying the goals outlined above, avoiding these problems, and
the conditions required for the method to work.

To set up a conservative lattice potential, many photons must be present in the
cavity field so that fluctuations can be neglected, and this necessitates strong
pumping from the in-coupled lasers. On the other hand, the probe field amplitude
must be small enough so that it does not affect the atomic dynamics, requiring
weak pumping. In addition, the probe and lattice fields couple to each other
through the scattering of photons off of the atoms. This acts as an extra source
for probe dynamics. The probe field is then not a direct measure of atomic
dynamics and can act back on the atoms, altering the properties we are
attempting to measure.

We can circumvent these problems by first choosing the relative phase on the in-coupled
lasers so that only one of two standing-wave modes in the cavity is pumped. Strong
pumping and the properties of a bad cavity -- where the fields are at all times in steady
state -- ensure that the pumped mode acts as a lattice potential. The other standing-wave
mode is not pumped. Any field leaking out of the cavity from this mode arises solely
because of events occurring in the cavity, and it can therefore act as a probe for system
dynamics.

Ensuring that the non-pumped mode acts as a probe of atomic dynamics requires that two
conditions be met. The probe must have as its source {\it only} the motion of the atoms.
This means that any probe dynamics due to the effective coupling to the lattice field must
be small compared to that induced by the motion of the atoms. The back-action of this
mode on the atoms must also be negligible. Any atomic motion induced by coupling
to the probe field must be small compared to the motion induced by both lattice and
external potentials.

Two conditions are also required for the pumped field to act as a conservative lattice
potential. The back-action of the atoms on the lattice field must be negligible, meaning
that deviations from the mean field amplitude caused by coupling to the atoms is small
compared to the mean field amplitude itself. In addition, any atomic motion induced by fluctuations away from the
mean lattice field must be small compared to that induced by the external potential,
since this is exactly the dynamics we want to measure.

The scheme is realized in the setup illustrated in Fig.
\ref{fig:systemschematic}. We consider $N_a$ ground-state atoms interacting with
two counter-propagating running-wave cavity modes in a ring resonator setup. The
two modes of the cavity have frequency, $\omega _{ca}$, and wave vectors,
$\pm k_{ca}\hat{z}$. The two cavity modes are coherently pumped at a detuning,
$\Delta_p=\omega_{L}-\omega_{ca}$, where $\omega_L$ is the frequency of the
pumping lasers. Photon decay through the cavity mirrors is treated within the
Born-Markov approximation. We treat the atom-cavity-field interaction in rotating
wave and dipole approximations. The cavity-modes are far-detuned from atomic
transitions.

\subsection{Model}

The effective Hamiltonian for the coupled atom-cavity system is given by
\begin{widetext}
\begin{eqnarray}\label{eqn:ham}
\hat{H} &=&\sum\int dz~\hat{\Psi}^{\dag}(z)\left(-\frac{\hbar^2}{2m}\frac{d^2}{dz^2}+V_{ext}(z)\right)\hat{\Psi}(z)
+\sum_{k=\pm k_{ca}}\left(\hbar\eta\hat{a}_{k}+\hbar\eta\hat{a}_{k}^{\dag}-\hbar\Delta_p\hat{a}_{k}^{\dag}\hat{a}_{k}\right) \nonumber\\
&&+\int dz~\hat{\Psi}^{\dag}(z)~\hbar g_0
\left(\hat{a}_{k_{ca}}^{\dag}e^{-ik_{ca}z}+\hat{a}_{-k_{ca}}^{\dag}e^{ik_{ca}z}\right)
\left(\hat{a}_{k_{ca}}e^{ik_{ca}z}+\hat{a}_{-k_{ca}}e^{-ik_{ca}z}\right)\hat{\Psi}(z).
\end{eqnarray}
\end{widetext}
Here, $\hat{\Psi}$ is the atomic field operator, and $\hat{a}_k$ is the annihilation operator
for the cavity mode, $k$. The parameter, $m$, is the the mass of the atom, $g_0$ is the
two-photon atom-cavity coupling, and $\eta$ is the strength of the cavity field pumping, taken to be real.
Due to the far detuning of the cavity fields from the atomic transition, excited states of the
atom have been adiabatically eliminated, and the atoms couple to the field intensity. The three terms
in Eq. (\ref{eqn:ham}) are respectively the atomic kinetic energy
and external potential, the bare cavity mode Hamiltonian, and the atom-cavity interaction. Cavity
losses through the cavity mirrors are treated via a master equation with Liouvillian,
\begin{equation}\label{eqn:Liouvillian}
\hat{\mathcal{L}}\hat{\rho}=-\frac{\hbar\kappa}{2}\sum_{k=\pm k_{ca}}\left(
\hat{a}_{k}^{\dag}\hat{a}_{k}\hat{\rho}+\hat{\rho}\hat{a}_{k}^{\dag}\hat
{a}_{k}-2\hat{a}_{k}\hat{\rho}\hat{a}_{k}^{\dag}\right),
\end{equation}
where $\hat{\rho}$ is the reduced density matrix for the atom-cavity system, and $\kappa$ is the cavity
linewidth.

We perform a canonical transformation of the cavity mode operators to symmetric
and anti-symmetric modes,
\begin{equation}\label{eqn:BpmDefs}
\hat{b}_{\pm}=\frac{\hat{a}_{k_{ca}}\pm\hat{a}_{-k_{ca}}}{\sqrt{2}}.
\end{equation}
The bare cavity Hamiltonian in terms of these operators is given by
\begin{equation}\label{eqn:BpmBareCavityHam}
\hat{H}_{ca}=\sqrt{2}\hbar\eta\left(\hat{b}_++\hat{b}_+^{\dag}\right)
-\hbar\Delta_p\left(\hat{b}_+^{\dag}\hat{b}_++\hat{b}_-^{\dag}\hat{b}_-\right).
\end{equation}
The symmetric mode, $\hat{b}_+$, is pumped by the in-coupled lasers whereas the anti-symmetric
mode, $\hat{b}_-$, is not. The $\hat{b}_+$ mode has a mode function proportional to
$\cos\left(k_{ca}z\right)$ and sets up the lattice potential, as follows.

The equation of motion for the symmetric field amplitude is
\begin{eqnarray}\label{eqn:BPlusEOM}
i\frac{d\langle\hat{b}_+\rangle}{dt}&=&\left(-i\frac{\kappa}{2}-\Delta_p\right)
\langle\hat{b}_+\rangle+\sqrt{2}\eta \nonumber\\
&&+~2g_0\int dz\cos^2\left(k_{ca}z\right)
\left\langle\hat{b}_+\hat{\Psi}^{\dag}\left(z\right)\hat{\Psi}\left(z\right)\right\rangle \nonumber\\
&&+~ig_0\int dz\sin\left(2k_{ca}z\right)\left\langle\hat{b}_-\hat{\Psi}^{\dag}(z)\hat{\Psi}(z)\right\rangle.
\end{eqnarray}
We perform another transformation to a fluctuation operator, $\hat{d}_{+}=\hat{b}_+-\beta$,
leaving the anti-symmetric mode unchanged, $\hat{d}_-=\hat{b}_-$. The mean steady-state
amplitude, $\beta$, is given by
\begin{equation}\label{eqn:BPlusSlaved}
\beta=\langle\hat{b}_+\rangle=\frac{\sqrt{2}\eta}{\Delta_p-2g_0\mathcal{C}(0)+i\kappa/2},
\end{equation}
where
\begin{equation}\label{eqn:C}
\mathcal{C}(t)=\int dz\cos^{2} \left(k_{ca}z\right)\left\langle\hat{\Psi}^{\dag}(z,t)\hat{\Psi}(z,t)\right\rangle,
\end{equation}

Assuming that the atom-field correlations factorize between atomic and field operators,
e.g. $\langle\hat{d}_{+}\hat{\Psi}^{\dag}\hat{\Psi}\rangle=
\langle\hat{d}_{+}\rangle\langle\hat{\Psi}^{\dag}\hat{\Psi}\rangle$, the equations of motion for both
$\hat{d}_+$ and $\hat{d}_-$ are given by
\begin{eqnarray}\label{eqn:bMinusEOM}
i\frac{d\langle \hat{d}_{-}\rangle }{dt}&=&\left(-\Delta_p+2g_{0}\mathcal{S}(t)-i\frac{\kappa}{2}\right)
\langle \hat{d}_{-}\rangle \nonumber \\
&&-~ig_{0}\left(\beta+\langle\hat{d}_+\rangle\right)\mathcal{S}_2(t),
\end{eqnarray}
and
\begin{eqnarray}\label{eqn:bFlucEOM}
i\frac{d\langle\hat{d}_+\rangle}{dt}&=&\left(-\Delta_p+2g_{0}\mathcal{C}(t)-i\frac{\kappa}{2}\right)
\langle\hat{d}_+\rangle\nonumber \\
&&+~ig_0\langle\hat{d}_-\rangle\mathcal{S}_2(t)+2g_0\beta\left(\mathcal{C}(t)-\mathcal{C}(0)\right),
\end{eqnarray}
where
\begin{eqnarray}
\mathcal{S}(t)&=&\int dz\sin^{2}\left(k_{ca}z\right)\left\langle\hat{\Psi}^{\dag}(z,t)\hat{\Psi}(z,t)\right\rangle,\label{eqn:S} \\
\mathcal{S}_2(t)&=&\int dz\sin\left(2k_{ca}z\right)\left\langle \hat{\Psi}^{\dag}(z,t)\hat{\Psi}(z,t)\right\rangle \label{eqn:S2}.
\end{eqnarray}
Finally, the equation of motion for the atomic field operator is given by
\begin{widetext}
\begin{eqnarray}\label{eqn:AtomEOM}
i\hbar\frac{d\hat{\Psi}(z)}{dt}&=&\left(-\frac{\hbar^2}{2m}\frac{d^2}{dz^{2}}+V_{lat}(z)
+V_{ext}(z)\right)\hat{\Psi}(z)+i\hbar g_0\left(\beta^{\ast}\hat{d}_--\beta\hat{d}_-^{\dag}
+\hat{d}_+^{\dag}\hat{d}_-
-\hat{d}_-^{\dag}\hat{d}_+\right) \sin\left(2k_{ca}z\right)  \hat{\Psi}(z) \nonumber \\
&&+~2\hbar g_0~\hat{d}_{-}^{\dag}\hat{d}_{-}\sin^{2}\left(k_{ca}z\right)\hat{\Psi}(z)+
2\hbar g_0\left(\hat{d}_+^{\dag}\hat{d}_+
+\beta\hat{d}_+^{\dag}+\beta^*\hat{d}_+\right)\cos^2\left(k_{ca}z\right)\hat{\Psi}(z),
\end{eqnarray}
\end{widetext}
where $V_{lat}(z)=V_0\cos^{2}\left(k_{ca}z\right)$ is a conservative
lattice potential of depth, $V_0=2\hbar g_0|\beta|^2$.

Aside from the conditions involving the external potential, the goals of
simultaneously setting up both an optical lattice potential and a weak
probe require that the inequalities,
\begin{equation}\label{eqn:AmplitudeIneqs}
|\langle\hat{d}_{\pm}\rangle|\ll|\beta|,
\end{equation}
are satisfied. This can be verified by examining equations
(\ref{eqn:bMinusEOM}), (\ref{eqn:bFlucEOM}), and (\ref{eqn:AtomEOM}).
These are necessary conditions, but finding the exact criteria
for neglecting the back-action requires a more careful
analysis of the problem, including numerical simulations. This is left for future work.
Ensuring that both the probe and the lattice fluctuations induce
atomic motion that is negligible compared to that induced by the
external potential requires explicit knowledge of the form of $V_{ext}$
and will therefore be left for the next section.

The two conditions in Eq. (\ref{eqn:AmplitudeIneqs}) can be be made more explicit. Equations
(\ref{eqn:bMinusEOM}) and (\ref{eqn:bFlucEOM}) imply the scaling relations,
\begin{eqnarray}
\langle\hat{d}_-\rangle&\sim&\frac{g_0\beta\mathcal{S}_2(t)}{\kappa}, \\
\langle\hat{d}_+\rangle&\sim&\frac{g_0\beta\left(\mathcal{C}(t)-\mathcal{C}(0)\right)}{\kappa},
\end{eqnarray}
so that the two conditions are respectively equivalent to
$|g_0\mathcal{S}_2(t)|\ll\kappa$ and $|g_0\left(\mathcal{C}(t)-\mathcal{C}(0)\right)|\ll\kappa$.
When these are satisfied, we may neglect Eq. (\ref{eqn:bFlucEOM}) altogether. In addition,
Eq. (\ref{eqn:bMinusEOM}) can be solved approximately since in this limit $\langle\hat{d}_-\rangle$
adiabatically follows the atomic motion. Finally, we have ensured that both fluctuations in the lattice
and the back-action of the probe field on the atoms can be neglected. We need then only keep the
first term in Eq. (\ref{eqn:AtomEOM}).

With these approximations in hand, the equation of motion for the atomic field operator is
\begin{equation}\label{eqn:AtomEOMSimple}
i\hbar\frac{d\hat{\Psi}(z)}{dt}=\left(-\frac{\hbar^2}{2m}\frac{d^2}{dz^{2}}+V_{lat}(z)
+V_{ext}(z)\right)\hat{\Psi}(z),
\end{equation}
and the probe field amplitude is given by
\begin{equation}\label{eqn:BMinusSlaved}
\langle\hat{d}_-(t)\rangle=\frac{-ig_0\beta}{\Delta_p-2g_0\mathcal{S}(t)+i\kappa/2}\mathcal{S}_2(t).
\end{equation}
These equations comprise a complete description of the coupled atom-cavity dynamics.

To the extent that the atoms are in the same center-of-mass state, $\ket{\psi(t)}$, satisfying Eq.
(\ref{eqn:AtomEOMSimple}), we can in Eq. (\ref{eqn:BMinusSlaved}) make the replacement,
\begin{equation}\label{eqn:sinMatElement}
\mathcal{S}_2(t)\to N_a \langle\psi(t)|\sin(2k_{ca}\hat{z})|\psi(t)\rangle.
\end{equation}

\subsection{Detection scheme}

\begin{figure}[tb]
\includegraphics[width=80mm]{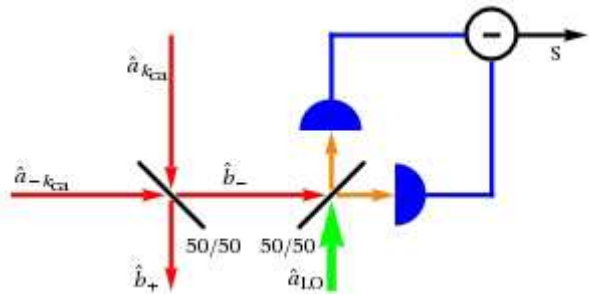}
\caption{(Color online.) Schematic of the balanced heterodyne detection scheme.
The out-coupled cavity beams, $\hat{a}_{\pm k_{ca}}$, are combined to form
symmetric ($\hat{b}_+$) and anti-symmetric ($\hat{b}_-$) modes. The antisymmetric
mode beats against a strong local oscillator (LO), $\hat{a}_{\rm LO}$ and photodetectors
count the number of photons in the quadratures of $\hat{b}_-$. The difference of these
counts is the signal.}
\label{fig:figureSchemeHet}
\end{figure}

Through Eq. (\ref{eqn:BMinusSlaved}), $\langle\hat{d}_-\rangle$ provides a measure
of the atomic dynamics within the cavity. In \cite{meiser2008sso}, two schemes for
detection of atomic motion using the out-coupled cavity fields were presented.
We here briefly review the superior case,
where heterodyne detection of $\hat{d}_-$ is performed by beating the field against a
strong local oscillator, as illustrated in Fig. \ref{fig:figureSchemeHet}.

According to input-output
theory~\cite{gardiner2004qn}, the field at the cavity output is proportional to
\begin{equation}
\hat{d}_{out}=\sqrt{\kappa}~{\hat{d}}_-+\hat{d}_{in}.
\end{equation}
By beating this field against a strong local oscillator, these photons can be detected with
unit efficiency. The input field state is the vacuum, in which case
$\langle\hat{d}_{out}\rangle=\sqrt{\kappa}~\langle\hat{d}_-\rangle$. The resulting signal
is the difference signal at the output of the photo-detectors, given by
\begin{equation}\label{eqn:Signal}
V(t)\propto{\rm Im}(\sqrt{\kappa}~\alpha_{\rm LO}^*\langle\hat{d}_-\rangle(t)),
\end{equation}
which is a product of $|\alpha_{\rm LO}|$ with
\begin{equation}\label{eqn:Quadrature}
\hat{q}_-=e^{-i\phi_{\rm LO}}\hat{d}_--e^{i\phi_{\rm LO}}\hat{d}_-^{\dag},
\end{equation}
a quadrature of the anti-symmetric mode field. The local oscillator amplitude is
$\alpha_{\rm LO}=|\alpha_{\rm LO}|e^{i\phi_{\rm LO}}$. The SNR is the ratio of the signal
power to signal variance, given by
\begin{equation}\label{eqn:SNR}
SNR=\int d\omega~|\sqrt{\kappa}~\langle{\hat{q}}_-\rangle(\omega)|^2.
\end{equation}
The integrand is proportional to the power spectrum, $S(\omega)$, of the signal current in
Eq. (\ref{eqn:Signal}).

\section{Results}\label{sec:Results}

In this paper, we consider the motion of atoms confined in the optical lattice in the
presence of gravity,
\begin{equation}
V_{ext}(z)=mgz,
\end{equation}
and use the scheme outlined in the previous section to probe the motion of the atoms.
Gravity measurements are important for instance for optical lattice clocks
\cite{ludlow2008slc}. For this reason, we treat the specific system of a gas of $^{87}$Sr
atoms, though the method certainly applies to many species of atoms. The
parameters for the coupled atom-cavity system are chosen to reflect current experimental
conditions. They are
$\lambda_{ca}=2\pi/k_{ca}=813~\rm{nm}$, $\hbar\kappa=100E_R$, $\Delta_p=0$,
$\hbar g_0=10^{-4}E_R$, and $N_a=10^4$, implying derived parameters of
$E_R\approx2\pi~4\rm~{kHz}~\hbar$ and $mgd\approx0.25E_R$;
$E_R=\hbar^2k_{ca}^2/2m$ is the recoil energy of the lattice, and $d=\pi/k_{ca}$ is
the lattice spacing.

We have to ensure that the back-action of both $\hat{d}_-$ and $\hat{d}_+$ on the atoms is still
negligible. Specifically, the coupling strengths in Eq. (\ref{eqn:AtomEOM}) must be small
compared to the characteristic coupling strength of $V_{ext}$, $\hbar\omega_B=mgd$.
These conditions are met if $|g_0\beta|^2\mathcal{S}_2(t)/\kappa\ll\omega_B$ and
$|g_0\beta|^2|\mathcal{C}(t)-\mathcal{C}(0)|/\kappa\ll\omega_B$. These inequalities are
well-satisfied for the parameters above. Again, while these conditions are necessary,
the exact criteria for being able to neglect the back-action of the fields on the atoms
requires more careful numerical study, which will be left for future work.

Within this setup, we envision an experiment in which the atoms are initially loaded
into a harmonic trap. A vertical one-dimensional optical lattice is slowly ramped on
so that the atoms are in the ground state of the combined potential of trap and lattice
for a non-interacting gas.
The trap is then switched off, and the gas is allowed to evolve under gravity. In the
presence of such a constant force, the atoms undergo Bloch oscillations. This
dynamics is briefly reviewed in the following discussion.

\subsection{System Dynamics}

\begin{figure}[tb]
\includegraphics[width=80mm]{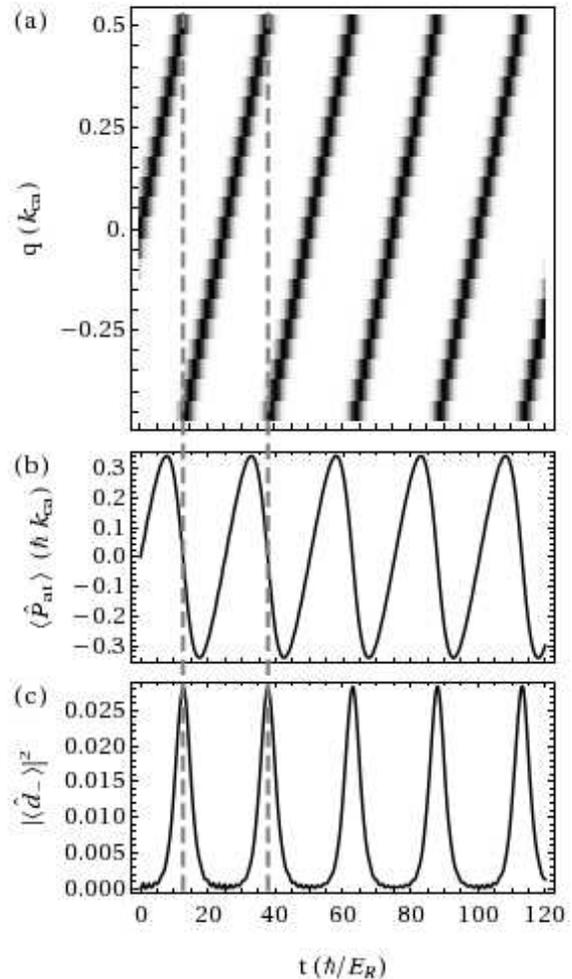}
\caption{Example of system dynamics for $V_0=-3E_R$ and initial
state a Gaussian of width $\sigma=2d$ projected into the first band. (a) Atomic
density in the first band plotted versus quasi-momentum. White corresponds to
zero population, black to maximal population. Population in the second band is at
most $0.001N_a$. (b) Expectation value of atomic momentum reflecting Bloch
oscillations. (c) Number of photons in the probe field.}
\label{fig:systemDynamics}
\end{figure}

The central result of the theory describing Bloch oscillations is based
on a semi-classical equation of motion~\cite{bendahan1996boa}, which states that the
average quasi-momentum of a wave-packet restricted to the first band increases linearly in
time until it reaches the Brillouin zone (BZ) boundary, at which point it is Bragg-reflected.
Explicitly, this is
\begin{equation}\label{eqn:AccTheorem}
\hbar\langle q \rangle(t)=\hbar\langle q\rangle(0)+mgt,
\end{equation}
where the quasi-momentum, $q$,  is restricted to the range, $-k_{ca}/2\leq q\leq k_{ca}/2$.
Since the group velocity of the atomic wave-packet is given by the derivative of the
dispersion relation~\cite{bendahan1996boa}, the periodic nature of the quasi-momentum
implies that the atomic momentum oscillates at a frequency, $\omega_B=mgd/\hbar$.
These Bloch oscillations will persist as long as there is negligible Landau-Zener tunneling
to higher bands. Each time the wave-packet reaches the BZ boundary, a fraction of
population is transferred to the second band, given by~\cite{morsch2001boa}
\begin{equation}\label{eqn:LZ}
P_{LZ}={\rm exp}\left(-\frac{\pi^2}{8}\frac{\Delta^2}{mgdE_R}\right),
\end{equation}
where $\Delta$ is the band-gap at the boundary. When $\Delta^2<4mgdE_R$, the
population transfer is appreciable, and vibrational dynamics significantly alter the behavior
of the atoms. For this reason, we restrict our attention to lattice depths greater than $3E_R$,
where $P_{LZ}$ is at most $10^{-5}$ for our choice of parameters.

In order to understand how Bloch oscillations are reflected in the time-dependence of the
probe field, we carefully consider Eq. (\ref{eqn:BMinusSlaved}). The operator,
$\sin(2k_{ca}\hat{z})$, is periodic in space with period $d$ and has odd parity, implying that
it connects two Bloch states, $\ket{\psi_q^{(n)}}$ and $\ket{\psi_{q'}^{(n')}}$, only
if the quasi-momenta are equal, $q=q'$, and the bands satisfy $n-n'={\rm odd}$. Taking
\begin{equation}\label{eqn:state}
\ket{\psi}=\sum_{n,q}c_q^{(n)}\ket{\psi_q^{(n)}},
\end{equation}
we can write the matrix element in Eq. (\ref{eqn:BMinusSlaved}) approximately as
\begin{equation}
\bra{\psi}\sin(2k_{ca}\hat{z})\ket{\psi}\approx \sum_q\rho_{q,q}^{(1,2)}+h.c.,
\end{equation}
where $\rho_{q,q}^{(1,2)}=c_q^{(1)*}c_q^{(2)}$ is the coherence between bands one and
two. This assumes an initial state confined to the first band in the case of negligible
coupling to bands three and higher. Using Eq. (\ref{eqn:AtomEOMSimple}), we can derive
an approximate equation of motion for the coherence; it is
\begin{equation}\label{eqn:CoherenceEOM}
i\frac{d\rho_{q,q}^{(1,2)}}{dt}=\Delta_q^{(1,2)}\rho_{q,q}^{(1,2)}+\omega_B\rho_{q}^{(1)},
\end{equation}
where $\rho_{q}^{(1)}$ is the population of the $q$-quasi-momentum state in the first band, and
$\hbar\Delta_q^{(1,2)}=E_q^{(2)}-E_q^{(1)}$ is the energy difference between the $q$-quasi-momentum
Bloch states in the first two bands. Since $\Delta_q^{(1,2)}\gg\omega_B$, the coherence follows the
first-band population adiabatically. In this approximation,
$\rho_{q,q}^{(1,2)}=-\rho^{(1)}_q\omega_B/\Delta_q^{(1,2)}$.

Combining this expression for a wave-packet that is narrow in quasi-momentum with Eq.
(\ref{eqn:AccTheorem}), Eq. (\ref{eqn:BMinusSlaved}) approximately becomes
\begin{equation}\label{eqn:DMinusApprox}
\langle\hat{d}_-(t)\rangle\approx\frac{ig_0\beta N_a}{\Delta_p-2g_0\mathcal{S}(t)+i\kappa/2}
\frac{\omega_B}{\Delta^{(1,2)}_{mgt}}.
\end{equation}
This expression implies that the probe field amplitude is largest when the atomic
wave-packet is centered at the BZ boundary since $\Delta^{(1,2)}_{mgt}$
is smallest at this point.

Equations (\ref{eqn:AtomEOMSimple}) and (\ref{eqn:BMinusSlaved}) are numerically
integrated for an initial state that is a Gaussian of spatial width, $\sigma$, projected
into the first band. This approximates the ground state of the combined potential of
lattice and harmonic trap for a non-interacting gas. An example of the system dynamics is illustrated in Fig.
\ref{fig:systemDynamics}, where $V_0=-3E_R$, and $\sigma=2d$. A vertical slice
through Fig. \ref{fig:systemDynamics}(a) is the wave function density in the first band
plotted versus quasi-momentum at an instant in time. The center of this wave-packet
moves linearly in time and is reflected at the BZ boundary ($q=k_{ca}/2$), as
in Eq. (\ref{eqn:AccTheorem}). Bloch oscillations are illustrated in
Fig. \ref{fig:systemDynamics}(b), where the atomic momentum oscillates in time.
Finally, the response of the probe field to this dynamics is illustrated in Fig.
\ref{fig:systemDynamics}(c). As predicted above, the probe field intensity peaks when
the atomic wave-packet reaches the BZ boundary.

\subsection{Signal and SNR}

\begin{figure}[tb]
\includegraphics[width=8cm]{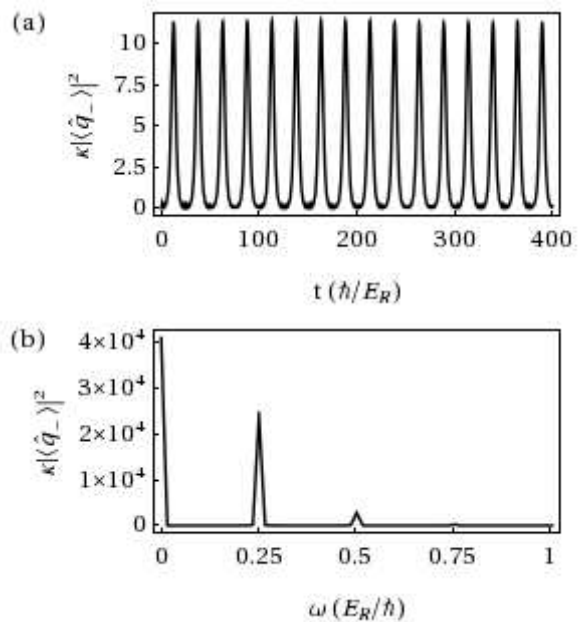}
\caption{Examples of the (a) signal power and (b) signal power spectrum, computed with parameters,
$V_0=-3E_R$, $\hbar g_0=10^{-4}E_R$, $N_a=10^4$, and $\hbar\kappa=100E_R$.
The signal displays a clear oscillation at the Bloch frequency, $\omega_B=0.25E_R/\hbar$.}
\label{fig:figureSignal}
\end{figure}

As described in Sec. \ref{sec:Model}, the probe field is combined at the cavity output with
a strong local oscillator, and the resulting signal is proportional to a quadrature of the
probe field, Eq. (\ref{eqn:Quadrature}). An example of such a signal is plotted in Fig.
\ref{fig:figureSignal}. There is a clear peak at the Bloch oscillation frequency in the signal
power spectrum, but there are also several harmonics present. In calculating the SNR, Eq.
(\ref{eqn:SNR}), we place a notch-filter about $\omega_B$ and count only the total
number of photons out-coupled from the quadrature at this frequency.

There are three properties of the system that can affect the SNR.
First, the width of the initial wave-packet has an effect. It is easiest to see why this is so by
taking as the initial state a Wannier function, which is a coherent superposition of Bloch
states in a single band, populated equally. According to Eq. (\ref{eqn:AccTheorem}), the
wave-packet is continuously reaching the BZ boundary, and the oscillation in the signal is
washed out. Second, when the lattice is too deep, the first two bands are essentially
flat, in which case $\Delta_q^{(1,2)}$ does not change with quasi-momentum,
eliminating the oscillations in the signal according to Eq. (\ref{eqn:DMinusApprox}).

The temperature of the atomic gas can also significantly influence the SNR. In a thermal cloud 
the replacement, Eq. (\ref{eqn:sinMatElement}), cannot be made, since the atoms do not all
occupy the same state. In this case, atoms in different lattice sites may contribute to the signal with
random phases, in which case the SNR scales with $N_a$ rather than $N_a^2$. The temperature
and chemical potential of the gas also determine the relative populations of the various Bloch states,
and appreciable population in higher bands can destroy Bloch oscillations. A proper treatment of
thermal effects is necessary for exact results, but here we assume the replacement, Eq.
(\ref{eqn:sinMatElement}), is a good approximation.

Equation (\ref{eqn:AtomEOM}) is numerically integrated for a time $t=400\hbar/E_R$.
The resulting wave function is used to compute the probe field amplitude,
Eq. (\ref{eqn:BMinusSlaved}), which is Fourier-transformed and squared, yielding the
power spectrum. The SNR is computed and scaled up linearly to an interrogation time
of $1$s, which assumes that coherence time of the Bloch oscillations is longer than $1$s.

\begin{figure}[tb]
\includegraphics[width=80mm]{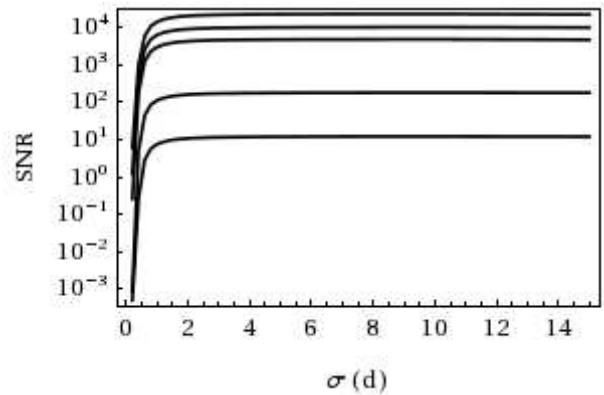}
\caption{(Color online.) Signal-to-noise ratio as a function of the initial wave-packet width for
an interrogation time of $1$s. The different plots correspond to lattice depths of (from largest
to smallest SNR) $V_0=-3,-4,-5,-10,-15E_R$.  For $\sigma<2d$, the SNR is reduced due to parts of
the wave-packet constantly moving past the Brillouin zone boundary, where the signal peaks. The
SNR saturates near $\sigma=2d$.}
\label{fig:figureSNR}
\end{figure}

\begin{figure}[tb]
\includegraphics[width=80mm]{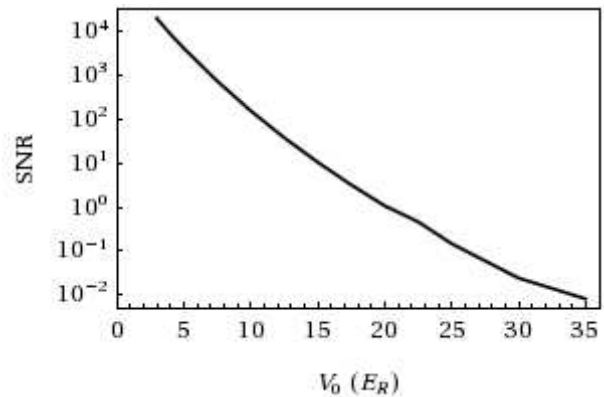}
\caption{Signal-to-noise ratio as a function of lattice depth for $\sigma=2d$ and an interrogation time
of $1s$. For $|V_0|<2E_R$, significant Landau-Zener tunneling to the second band destroys Bloch
oscillations. For $|V_0|>3E_R$, the SNR decreases due to the increasing flatness of the lowest band,
which in turn decreases the amplitude in momentum space of the Bloch oscillations.} 
\label{fig:figureSNR2}
\end{figure}

The results are plotted in Figs. \ref{fig:figureSNR} and \ref{fig:figureSNR2}. The SNR climbs from zero
for small wave-packet widths and saturates near $\sigma=2d$. The decrease in SNR for $\sigma<2d$
is a result of the fact that the wave-packet is wide in quasi-momentum, which means that a significant
portion of the wave-packet is at the the BZ boundary for all times. We get a maximum when the lattice
depth is relatively small, $|V_0|\approx3E_R$, and the SNR decreases with increasing lattice depth.


\section{Conclusion}\label{sec:Conclusion}

We have described a general cavity QED system in which properties of
atomic dynamics can be probed {\it in situ} and non-destructively. One
cavity field is strong enough to act as a conservative lattice potential for
the atoms, and the other cavity field is weak so that it acts as a
non-destructive probe of atomic motion. This technique is applied to
the detection of Bloch oscillations. Balanced heterodyne detection of
the probe field at the cavity output combined with integration in time
and across the atomic cloud allows for SNRs as high as $10^4$. 

Examining Eqs. (\ref{eqn:BMinusSlaved}) and (\ref{eqn:SNR}), we can see that
the SNR can be increased by either decreasing the cavity linewidth, $\kappa$, at
fixed lattice depth and atom-cavity coupling or increasing the coupling constant,
$g_0$, at fixed $V_0$ and $\kappa$. The linewidth can be increased as long as
the system remains in the bad cavity limit. However, a linewidth of
$\kappa=100E_R$ is already very small from an experimental standpoint, so
increasing it beyond this level is a technological challenge. On the other hand,
$g_0$ can be varied merely by varying the detuning between the cavity fields
and atomic transitions. In addition, the SNR scales with the square of the
number of atoms, so increasing $N_a$ beyond the $10^4$ level assumed in this
paper is also desirable. This can all be done to the extent that the conditions
outlined in Sec. \ref{sec:Model} and Sec. \ref{sec:Results} are still met. 

This scheme can be extended for use in detection of various atomic
properties, and the measurement of Bloch oscillations itself can be viewed
as a general DC probe for atomic dynamics and states. For instance, Bloch
oscillations may be used for measurement of fundamental constants
\cite{clade2006dfs} and for Casimir forces~\cite{wolf2007fol}.
Varying the detuning between two lattice beams gives rise to an effective
acceleration of the lattice~\cite{morsch2001boa}, and band physics may
be probed by varying the Bloch oscillation frequency in such a setup.
Breakdown of Bloch oscillations are a signature of many-body
effects in an atomic gas~\cite{freericks2008qbo}, and this is
signalled by a reduction in SNR compared to the non-interacting case.

Two generalizations of this measurement technique are readily realizable.
We may implement a periodic forcing whose varying driving frequency
can be a spectroscopic probe of atomic dynamics. The simplest examples of this
include shaking the lattice~\cite{ivanov2008cda} and modulating the amplitude
of the lattice~\cite{alberti2009cha}. Another important extension of the method
involves measuring higher-order correlation functions of the out-coupled probe
field. Since one cavity field operator couples to two atomic field operators (see
for instance Eq. (\ref{eqn:bMinusEOM})), higher-order properties of the atoms
such as density-density correlations can easily be measured with standard
quantum optical techniques. The use of higher-order correlation functions of the
probe field is a starting point for generalizing this technique to probe many-body
physics in optical lattices.

We acknowledge useful conversations with Jun Ye, Ana Maria Rey, and Victor Gurarie. This work was
supported by DARPA, NIST, DOE, NSF, DFG (DM), and ASI grant WP4200 (MC).

\bibliographystyle{apsrev}
\bibliography{BlochOscillations}

\end{document}